%% file: template.tex
\documentclass{Interspeech}

\interspeechcameraready

\title{On-device Streaming Discrete Speech Units}

\author[equalcontribution]{Kwanghee}{Choi}
\author[equalcontribution]{Masao}{Someki}
\author[]{Emma}{Strubell}
\author[]{Shinji}{Watanabe}

\affiliation{Language Technologies Institute}{Carnegie Mellon University}{USA}
\email{\{kwanghec,msomeki,estrubel,swatanab\}@andrew.cmu.edu}

\usepackage{comment}
\usepackage{subfig}
\usepackage{cleveref}
\usepackage{cite}
\usepackage{multirow}
\usepackage{tabularx, booktabs}

\begin{document}

\maketitle
% \ifinterspeechfinal
% \thx{* Equal contribution.}
% \fi
% https://docs.google.com/spreadsheets/d/1WzIFS-5jOIi82x75C8hcbBzUraa8BqCaM8EOOTR7iN4/edit?gid=0#gid=0

% WAVLab template: \url{https://www.overleaf.com/read/kxqzhvqwdnjr#555764}

\begin{abstract}
% Recent advancements in speech tokenization have leveraged discrete speech units (DSUs) derived from clustering the features of self-supervised speech models (S3Ms).
Discrete speech units (DSUs) are derived from clustering the features of self-supervised speech models (S3Ms).
DSUs offer significant advantages for on-device streaming speech applications due to their rich phonetic information, high transmission efficiency, and seamless integration with large language models.
However, conventional DSU-based approaches are impractical as they require full-length speech input and computationally expensive S3Ms.
In this work, we reduce both the attention window and the model size while preserving the effectiveness of DSUs.
Our results demonstrate that we can reduce floating-point operations (FLOPs) by 50\% with only a relative increase of 6.5\% in character error rate (CER) on the ML-SUPERB 1h dataset.
These findings highlight the potential of DSUs for real-time speech processing in resource-constrained environments.
\end{abstract}

\section{Introduction}
Self-supervised speech models (S3Ms) have demonstrated remarkable efficacy in various speech-related tasks, including automatic speech recognition (ASR), text-to-speech synthesis, speech translation, speaker identification, and emotion recognition~\cite{mohamed2022self,wav2vec2,chen2022wavlm}.
With the advent of S3Ms, many have started to leverage their features for tokenizing speech, \textit{i.e.}, producing discrete speech units (DSUs)~\cite{chang23b_interspeech,chang2024exploring,chang24b_interspeech,borsos2023audiolm,SpeechGPT,rubenstein2023audiopalm,nakamura2025discrete}.
DSUs are typically generated by applying clustering methods such as k-means to S3M features, with the resulting clusters treated as DSUs.

The characteristics of DSUs are well adapted to on-device streaming scenarios, offering cheap transmission and easy adaptation to existing large language models (LLMs).
Unlike high-dimensional speech features, such as spectral speech features or S3M features, DSUs offer improved data storage and transmission efficiency~\cite{chang23b_interspeech,nakamura2025discrete} while maintaining comparable ASR performance~\cite{chang2024exploring}.
These advantages make DSUs a promising medium for speech data transmission.
For instance, a 1-second 16kHz audio signal requires 512kbps in raw form.
S3M features (\textit{e.g.}, wav2vec 2.0-large~\cite{wav2vec2}) increase this to 1600kbps.
However, DSUs with 2k clusters reduce this to just 0.6kbps, significantly compressing the data by 3-4 orders of magnitude.
DSUs also offer a promising way to extend LLMs to speech modalities~\cite{SpeechGPT,anygpt,rubenstein2023audiopalm}.
A practical application involves generating DSUs on client devices and transmitting them to server-side LLMs for processing.
This may also provide benefits for anonymizing personally identifiable information for privacy concerns \cite{hernandez2024anonymizing} as S3Ms lose speaker information at later layers \cite{choi2024self}.

\input{figures/fig_pareto}
However, DSUs have yet to be leveraged for streaming on edge devices due to two major challenges.
First, conventional S3Ms rely on bidirectional Transformer encoder architectures, with full attention window size to cover the entire input speech, which is not immediately compatible with the streaming setting.
Second, state-of-the-art S3Ms typically require hundreds of millions of parameters, which can be computationally infeasible to execute on edge devices with limited computational capacity.
It poses a problem for on-device streaming, where GPU memory restricts the total sequence length and prevents real-time generation of DSUs.%, as the model must process the entire speech input before generating DSUs.
% The second challenge comes from the fact that S3Ms typically have hundreds of millions of parameters.
% In on-device scenarios, it becomes computationally infeasible.
% In summary, for on-device streaming scenarios, extracting DSUs should have a limited amount of future frames while being computationally lightweight.

In summary, a model for extracting DSUs must work well even with a limited window of future frames while being computationally lightweight.
Evidence from previous work suggests this should be possible.
For example, prior work has indicated that S3Ms predominantly encode phonetic information, with limited syntactic and semantic content~\cite{pasad2023comparative,choi2024self,choi2025leveraging,choi2024understanding}.
Furthermore, studies have shown a strong correlation between extracted DSUs and phonemes~\cite{wav2vec2,hsu2021hubert}.
These findings indicate that DSUs can be extracted with a smaller window size.
Regarding the model size, we hypothesize that we can reduce the number of required parameters by solely targeting the specific input distribution, similar to the ideas of previous work \cite{lai2021parp,peng2023structured}.
Typical S3Ms are self-supervised on diverse data to perform well on a wide variety of audio, \textit{i.e.}, ``universal audio feature.''
However, in practice, DSUs would still be useful even if tailored for specific target data and tasks.

In this work, we investigate the feasibility of DSUs for streaming on edge devices by exploring the following two key questions: 1. How much temporal window size is needed? (\Cref{sec:width}); 2. How much model capacity (parameters) is needed? (\Cref{sec:depth}).
We investigate tradeoffs between computational efficiency and downstream performance in this setting by varying the attention window size and the number of layers of WavLM-large~\cite{chen2022wavlm}, a representative S3M for extracting DSUs~\cite{chang24b_interspeech}.
For each setting, a smaller model is trained by regarding the full model's DSUs as ground truth.
We evaluate feasibility by measuring the computational cost, comparing with ASR performance on Librispeech \cite{panayotov2015librispeech} and ML-SUPERB \cite{ml_superb}.

Our contributions in this paper are summarized as follows:

\begin{itemize}
    \item To the best of our knowledge, this study is the first to explore DSUs in the on-device streaming setting, exploring the feasibility of real-time lightweight computation of DSUs.
    \item We explore the trade-off between downstream performance and computational overhead by varying the attention window size and the number of layers of the original S3M, obtaining Pareto optimal trade-off curve.
    \item By optimizing the model's attention window size and number of layers, we reduce FLOPs by 50\% while having only a 6\% relative increase in CER on the ML-SUPERB 1h dataset.
\end{itemize}

\section{Experimental Settings}
Our work investigates whether DSUs can be accurately predicted using a smaller model.
To evaluate the predicted DSUs, we focus on the discrete ASR system.
We use the same experimental settings across all experiments.

\subsection{Discrete ASR System}
Discrete ASR system contains two main modules: speech-to-unit (S2U) module and unit-to-text (U2T) module.
S2U module generates DSUs, \textit{i.e.}, speech discretization from 
continuous S3M features.
Given the input raw speech $\mathbf{x} \in \mathbb{R}^{t}$ with length $t$, $f_\text{S3M}$ transforms speech into S3M features:
\begin{align}\label{eq:s3m}
    \mathbf{S} = f_\text{S3M}(\mathbf{x}) \in \mathbb{R}^{T \times F},
\end{align}
where $F$ is the feature dimension and $T$ is the temporal dimension roughly proportional to $t$.
Then, the pretrained k-means algorithm is applied to yield the stream of DSUs $\mathbf{D} = f_\text{kmeans}(\mathbf{S})$.
The cascade of S3M and k-means becomes the S2U module: \begin{align}\label{eq:s2u}
    f_\text{S2U}(\mathbf{x}) = f_\text{kmeans}(f_\text{S3M}(\mathbf{x})) = \mathbf{D} = [D_1, D_2, \cdots, D_T],
\end{align}
where $D_i \in \{1, 2, \cdots, V\}$ is the DSU at timestep $i$ and $V$ is the vocabulary size, \textit{i.e.}, $k$ in k-means.

After generating DSUs from speech through the S2U module (\cref{eq:s2u}), U2T module translates DSUs to text transcriptions.
To reduce the sequence length, additional post-processing methods are often applied, such as deduplication or byte-pair encoding tokenization \cite{chang2024exploring,chang24b_interspeech}:
\begin{align}\label{eq:subword}
    \mathbf{D}' = f_\text{subword}(\mathbf{D}) = [D'_1, D'_2, \cdots, D'_{T'}],
\end{align}
where $D'_j \in \{1, 2, \cdots, V'\}$ is the post-processed DSU at timestep $j$.
Often, the sequence length $T'$ becomes shorter and vocabulary size $V'$ becomes larger, \textit{i.e.}, $T' \leq T$, and $V' \geq V$.

Finally, the discrete ASR model $f_\text{ASR}$ is trained to predict the transcription $\mathbf{y}$.
The cascade of post-processing methods (\cref{eq:subword}) and discrete ASR model becomes the U2T module:
\begin{align}\label{eq:u2t}
    f_\text{U2T} (\mathbf{D}) = f_\text{ASR}(f_\text{subword}(\mathbf{D})) = f_\text{ASR}(\mathbf{D}') = \hat{\mathbf{y}},
\end{align}
where $\hat{\mathbf{y}}$ is the prediction of $\mathbf{y}$.

In summary, transcription $\mathbf{y}$ is predicted from input speech $\mathbf{x}$ through the cascade of S2U (\cref{eq:s2u}) and U2T (\cref{eq:u2t}):
\begin{align}\label{eq:e2e}
    \hat{\mathbf{y}} = f_\text{U2T}(f_\text{S2U}(\mathbf{x})) = f_\text{ASR}(f_\text{subword}(f_\text{kmeans}(f_\text{S3M}(\mathbf{x})))),
\end{align}
where the DSUs $\mathbf{D} = f_\text{kmeans}(f_\text{S3M}(\mathbf{x}))$ becomes the communication medium between the two modules.
Often, both modules are not trained in an end-to-end manner, but separately trained.

\subsection{Training a DSU predictor}
Our aim is to train a lightweight DSU predictor $\bar{f}_\text{S2U}$ that predicts the DSUs $\mathbf{D}$ given the speech input $\mathbf{x}$, so that it can replace $f_\text{S2U}$ within the discrete ASR system.
To make $\bar{f}_\text{S2U}$ to be trainable in an end-to-end manner, we replace the k-means model $d$ with a trainable fully-connected layer (FC):
\begin{align}\label{eq:dsu_logits}
    f_\text{FC}(f_\text{S3M}(\mathbf{x})) = [\hat{\mathbf{D}}_1, \hat{\mathbf{D}}_2, \cdots, \hat{\mathbf{D}}_T],
\end{align}
where $\hat{\mathbf{D}}_i \in \mathbb{R}^V$ is the pre-softmax logits at timestep $i$.
The final predicted DSU becomes $\hat{D}_i = \text{argmax}_v \hat{\mathbf{D}_i}[v]$, such that:
\begin{align}\label{eq:dsu_pred}
    \bar{f}_\text{S2U}(\mathbf{x}) = \text{argmax}_v f_\text{FC}(f_\text{S3M}(\mathbf{x})) = [\hat{D}_1, \hat{D}_2, \cdots, \hat{D}_T],
\end{align}
with a slight abuse of notation for the $\text{argmax}$ operator.
In summary, we use the original DSUs $\mathbf{D}$ as the ground truth label for the lightweight DSU predictor $\bar{f}_\text{S2U}$, so that we can replace the original S2U module $f_\text{S2U}$ within the discrete ASR system.

\subsection{System Evaluation}\label{ssec:setting_eval}
We use the existing discrete ASR system challenge~\cite{chang24b_interspeech}, which uses frozen WavLM-Large~\cite{chen2022wavlm} 21st layers' features to extract DSUs.
The challenge uses two datasets: LibriSpeech~\cite{panayotov2015librispeech} and ML-SUPERB 1-hour subset~\cite{ml_superb}, which covers English and 143 languages, respectively.
We denote the clean and noisy subset of LibriSpeech as test-clean and test-other, and ML-SUPERB test set as test-1h.
To measure the downstream performance, we use word error rate (WER) for test-clean and test-other, and character error rate (CER) for test-1h, following~\cite{chang2024exploring}.
Also, we use tera ($10^{12}$) floating point operations (TFLOPs) per one minute of input audio to measure the computational cost of the S2U module (\cref{eq:s2u,eq:dsu_pred}).
It differs from the original challenge, which uses bitrate to focus on the transmission efficiency of the DSU itself, not the overhead of extracting DSUs.
We use calflops \cite{ye2023calflops} to calculate FLOPs.
% For both cost and latency, we only measure the overhead of the S2U module (\cref{eq:s2u,eq:dsu_pred}).

\subsection{Experimental Details}\label{ssec:setting_detail}
Following~\cite{chang24b_interspeech}, we use the 12-layer E-branchformer encoder~\cite{kim2023branchformer} and 6-layer Transformer decoder~\cite{waswani2017attention} for the discrete ASR model $f_\text{ASR}$ (\cref{eq:u2t}).
However, for computational efficiency, we reduce the beam size from 20 to 5.
For all of our experiments, we used the AdamW optimizer~\cite{loshchilov2018decoupled} with learning rate 1e-4 and weight decay 1e-6.
We used the step learning rate decay with a rate of 0.9 every 1K steps.
We use the variable batch size with 2M frames for 10 epochs, setting patience to 1.
Refer to our codebase for more details.\footnote{\url{https://github.com/Masao-Someki/StreamingDSU}}

\vspace{1em}

\input{figures/fig_window}
\section{Reducing the Attention Window Size}\label{sec:width}
\textbf{Why does window size matter?}
S3Ms leverage Transformers with the full attention window, which requires the full audio to produce discrete units, making the streaming scenario impossible.
% For example, in principle, to calculate the DSU with the current structure, S2U model requires full speech.
However, this issue can be mitigated by limiting the future window size of the S2U module.
% For example, if the model has to observe a 1-second amount of future frames before predicting the DSU, it means that the theoretical latency becomes 1 second.
The smaller the required future frames, the smaller the theoretical delay of the system for streaming scenarios.
Additionally, reducing the input size enables lower computational overhead. %, as the Transformer's complexity is $O(n^2)$ with sequence length $n$.
% For example, extracting DSUs from a 1-hour-long recording becomes computationally infeasible.
% On the other hand, the problem can be avoided if the model can successfully predict DSU with a fixed window size.

\textbf{Experimental settings.}
We aim to answer whether the entire input is required or if we can yield similar downstream performance with limited attention window size.
We regard the case where we input the full speech as the strong baseline (\cref{eq:s2u}).
We test various window sizes by varying the number of past and future frames.
Motivated by the codebase of WavLM~\cite{chen2022wavlm}, we use a streaming mask, \textit{i.e.}, time-restricted self-attention~\cite{povey2018time,moritz2020streaming}, to limit the window size.
We fully fine-tune the DSU predictor $\bar{f}_\text{S2U}$ (\cref{eq:dsu_pred}) while using the same U2T module $f_\text{U2T}$ (\cref{eq:u2t}) as the baseline.

% We use the case where we use no past and future frame as the upper baseline for CER/WER, i.e., using only one frame.
We denote various attention window configurations as the number of left, center, and right frames, \textit{i.e.}, $[l, c, r]$.
Given the number of layers $n$, the temporal receptive field size is $(l+r)\times n + c$ and the theoretical latency is $r\times n+c$.

We always set the center frame as $c=1$,%, targeting frame-by-frame streaming systems with limited future context to control latency, computational cost, and ASR performance.
and vary the number of left and right frames, \textit{i.e.}, $l$ and $r$.
We first test the symmetric windows: $l=r=1, 2, \cdots, 64$.
% Maximum is set to 64, where 129 frames ($[l=64, c=1, r=64]$) translate to 2.6 seconds, roughly covering a full sentence.
Additionally, we aim to measure the usefulness of the past or future frames.
As such, we compare past-only ($[l=1, 2, \cdots, 128, c=1, r=0]$) and future-only ($[l=0, c=1, r=1, 2, \cdots, 128]$) windows.

\textbf{Results.}
The results in \Cref{fig:window} align with expectations, showing that performance improves as the attention window size increases.
However, the magnitude of improvement diminishes with progressively larger window sizes.
It suggests that DSUs capture contextual information from surrounding frames~\cite{choi2025leveraging}.
Additionally, even with the same window size, models that consider only past or future frames perform significantly worse than those that incorporate both.
In the case of test-clean, future frames contribute slightly more to performance than past frames, whereas test-other and test-1h show little to no difference, supporting the effectiveness of a symmetric window approach.
The results suggest there is little distinction between using past and future information.
% However, we leave a more detailed exploration for future work.
Further, we explore additional techniques to improve performance in \Cref{sec:pareto}.

% \blue{if we have time, add expr on removing pos encoding case?}

\section{Reducing the Number of Layers}\label{sec:depth}
\input{figures/fig_baseline_comparison}
\input{figures/fig_layer}
\input{figures/fig_sota_window}
\textbf{Why does the number of layers matter?}
Existing S3Ms typically consist of a large number of layers.
Since the number of layers has a linear relationship with computational cost, reducing them benefits resource-constrained on-device applications.

Also, through this experiment, we aim to estimate the amount of compute required to achieve a certain downstream performance.
As it is empirically known that neighboring layers tend to contain similar amounts of information~\cite{pasad2023comparative,choi2024self,choi2025leveraging}, fine-tuning S3Ms while removing the final layer one by one is the most straightforward approach.

% S3Ms are trained in a self-supervised manner without the explicit consideration of extracting DSUs.
% In other words, DSUs are a side-product of learning general speech representations.
% Furthermore, our setting focuses on the specific data distribution, not requiring the representational power for general speech and audio.
% Hence, a natural question arises: \textit{Do we need all the S3M's Transformer layers to extract the DSUs?}

\textbf{Experimental settings.}
To answer the question, we designed a simple experimental setting of reducing Transformer layers one by one and extracting DSUs from them.
For $f_\text{S3M}$ (\cref{eq:s3m}), we use layers up to layer index 21, 18, $\cdots$, 3, 0, where 0 means using only convolutional features.
For the DSU predictor $f_\text{S2U}$ (\cref{eq:dsu_pred}), we compare the case where we only fine-tune the FC layer $f_\text{FC}$ (DSU Frozen) and also fine-tuning the S3M layers $f_\text{S3M}$ (DSU FT).
We use the frozen U2T module $f_\text{U2T}$ from the challenge baseline.
We additionally compare the case where we also fine-tune the U2T module $f_\text{U2T}$ (DSU Full FT).

We compare with three baselines, where it is summarized in \Cref{fig:baseline_comp}: (1) the original challenge baseline (\cref{eq:s2u}, DSU Baseline); (2) using the frozen S3M features directly (SSL Frozen); and (3) also fine-tuning the S3M features (SSL FT).
% training an encoder-decoder model that takes as input frozen S3M's k-th layer features and outputs text (SSL); and (3) training both encoder-decoder model and S3M with the k-th layer output (SSL FT).
% B1 becomes the strong baseline for DSU-based models, as we consider them the ground truth for the DSU estimator and use the frozen unit-to-text model.
SSL Frozen is the strong baseline for DSU Frozen, which contains the full information inside the frozen $n$-th layer features.
% B3 is included to compare S3M representations with traditional filterbanks.
Similarly, SSL FT is the strong baseline for DSU FT, which demonstrates the case when we fully utilize the available model weights.

For SSL FT and SSL Frozen, we follow the settings of \cite{chang2024exploring}, which slightly modifies the U2T module $f_\text{U2T}$.
We feed S3M features to the linear layer and feed through the Transformer directly.
We denote the module $f_\text{F2T}$, \textit{i.e.}, feature-to-text module.
We use the same training settings as \Cref{sec:width} except for the learning rate scheduler for SSL Frozen and SSL FT; a warmup of 15k was empirically necessary for convergence.
% Also, to feed the continuous features into the same encoder-decoder architecture, we use a learnable linear layer that projects to 128-dimensional feature.

\textbf{Results.}
Similar to \Cref{sec:width}, an expected trend is observed in \Cref{fig:layer}: reducing the number of layers leads to performance degradation and fine-tuning more modules leads to better performance.
In the lowest layers, SSL features generally perform well, suggesting that relevant information is present but largely lost during the DSU prediction.
Nonetheless, DSU-based methods remain generally on par with SSL approaches.
In particular, SSL Frozen and SSL FT tend to overfit more easily to the majority language (test-clean and test-other), resulting in degraded performance on other languages (test-1h).
% Notably, SSL performs worse on test-1h than test-clean and test-other, indicating potential overfitting to the high-resource language (English of Librispeech).
% This effect is likely amplified by the fine-tuning of the feature-to-text model.

\section{Towards the Pareto Optimal}\label{sec:pareto}
We test various methods to improve downstream performance.
Also, by applying such, we produce a Pareto optimal curve that represents the trade-off between the computational overhead and the downstream performance.

\textbf{Learnable weights for multi-layer features (WF).} \label{subsec:weighted_feats}
As demonstrated in~\cite{pasad2023comparative,choi2024self}, different layers encode different types of information.
To leverage this, prior work~\cite{superb,ml_superb,chen2022wavlm} has employed a learnable weighted summation of features across layers, removing the need to search for the optimal layer.
Motivated by this approach, we apply the same technique to extract S3M features for DSU prediction.

\textbf{Fine-tuning unit-to-text module (U2T FT).}\label{subsec:finetune_u2t}
We hypothesize that DSU predictors with smaller window (\cref{eq:dsu_pred}) will produce noisier DSU predictions, where the U2T module (\cref{eq:u2t}) has not been exposed to during training.
To address this, we fine-tune the U2T module to enhance its robustness against such noise, resulting in better performance in downstream tasks.

\textbf{Experimental settings.}
For WF, we use the exact same training settings as before.
For U2T FT, we use the same settings except for the variable batch size of 20K and patience 3.
We choose two baselines: (1) the original challenge baseline; (2) limiting only the future window size $r$, \textit{i.e.}, $[l=\infty, c=1, r=1,2,\cdots,64]$ (Full Past).
We show all methods' future attention window size for extracting DSUs.
The theoretical latency, which is proportional to the future window size, is often used in streaming scenarios \cite{fu2024wav2vec}.
Note that symmetric windows' computational cost scale linearly with speech length, whereas both baselines scale quadratically.

% We use the same setting from \Cref{sec:width}.

\textbf{Results.}
\Cref{fig:sota-window} demonstrates that Full Past outperforms Symmetric for smaller windows.
However, the difference diminishes beyond 0.5s window size, suggesting a fixed past window with an adjustable future window for latency needs.
For WF, it reduces the largest windows' performance drop from 57\% and 83\% to 18\% and 40\% in test-clean and test-other.
However, it shows limited improvement on ML-SUPERB.
On the other hand, U2T FT is effective, especially for smaller windows across all datasets, supporting the above hypothesis that fine-tuning may learn noise in smaller window predictions.

\textbf{Pareto optimal curve.}
By integrating all findings and techniques, we construct the Pareto optimal curve in \Cref{fig:pareto}.
We use symmetric windows of varying sizes ($l=r=1,2,\cdots,64, c=1$) while varying the number of layers (12, 15, 18, 21).
We also apply learnable weights and test both with and without U2T fine-tuning.

Our Pareto curve is shown in \Cref{fig:pareto}, demonstrating that one can extract DSUs in a lightweight streaming way with reasonable trade-off on ASR performance.
Especially with 0.96 TFLOPs, we achieve 6.3 WER (test-clean), 14.0 WER (test-other), and 24.4 CER (test-1h), close to the baseline performance of 5.0 WER, 9.2 WER, and 23.0 CER.

\section{Related Works}

\textbf{Making S3Ms lightweight.}
Knowledge distillation (KD) is often used to reduce the size of S3Ms, such as DistilHuBERT \cite{chang2022distilhubert} or FitHuBERT \cite{lee2022fithubert}.
Techniques like pruning \cite{lai2021parp,peng2023structured} and quantization \cite{oujabour2025adaptive} is also used.
These approaches, evaluated on general benchmarks \cite{ml_superb}, primarily aim to create generally usable S3Ms with lower computational cost, often maintaining the non-streamable Transformer architecture.
Our work focuses on efficient streaming DSUs, fine-tuning the model for a specific task and data, rather than yielding general-purpose features.

\textbf{Speech tokenization.}
Speech tokenization broadly falls into acoustic tokens \cite{zeghidour2021soundstream,fossez2023high,shi2024espnet} or DSUs based on S3Ms \cite{wav2vec2,chen2022wavlm,hsu2021hubert}.
Acoustic tokenizers are often trained with the Vector Quantised-Variational AutoEncoder (VQ-VAE) \cite{van2017neural}), which emphasizes real-time, high-fidelity acoustic detail.
On the other hand, DSUs are known to contain rich phonetic information \cite{pasad2023comparative,choi2024self,choi2025leveraging,choi2024understanding} and hence provide structural coherence \cite{borsos2023audiolm} and better ASR performance~\cite{chang2024exploring}, although often offline.
Rather than relying on VQ-VAE, our work aims to make DSUs directly streamable and lightweight, maintaining their rich knowledge while enabling real-time abilities.

\textbf{DSUs for KD.}
% KD using DSUs appears across diverse speech subfields.
\cite{OliveiraG24} uses DSUs to distill S3Ms, but not specifically optimizing for predicting DSUs.
DSUs have also been used for KD in voice conversion \cite{kanagawa2024knowledge} or to guide acoustic tokenizers such as Speechtokenizer \cite{zhang2024speechtokenizer} and Mimi \cite{defossez2024moshi}.
However, these approaches either prioritize building general speech models or introduce complex multi-stage systems.
In contrast, our work directly concentrates on enhancing the efficiency of the DSU themselves, specifically for on-device streaming cases.

\section{Conclusion}
DSUs offer advantages for on-device streaming due to their transmission efficiency and compatibility with LLMs.
However, current methods for generating DSUs rely on full speech input and computationally heavy S3Ms.
Therefore, we investigated both the attention window size and the number of layers in S3Ms.
Our experiments demonstrate the feasibility of creating lightweight and streamable DSUs.
Furthermore, we show that weighted feature summation and fine-tuning the unit-to-text model effectively improve performance.
Finally, we explored the trade-off between ASR performance and computational overhead, establishing Pareto optimal curve.

\ifinterspeechfinal
\section{Acknowledgements}
This study was supported by the BRIDGE program of the Cabinet Office, Government of Japan.\hspace{-0.3em}
We also used the Bridges2 system at PSC and Delta system at NCSA through allocations CIS210014 and IRI120008P from the Advanced Cyberinfrastructure Coordination Ecosystem: Services \& Support (ACCESS) program, through National Science Foundation grants \#2138259, \#2138286, \#2138307, \#2137603, and \#2138296.
\else
\fi

\bibliographystyle{IEEEtran}
\bibliography{mybib}

\newpage
\input{figures/appendix_tables}

\end{document}

%% file: figures/fig_pareto.tex
\begin{figure}[t]
    \centering
    \subfloat{\includegraphics[width=0.15\textwidth]{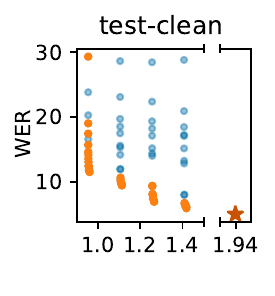}}
    \subfloat{\includegraphics[width=0.15\textwidth]{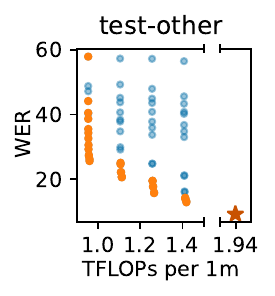}}
    \subfloat{\includegraphics[width=0.15\textwidth]{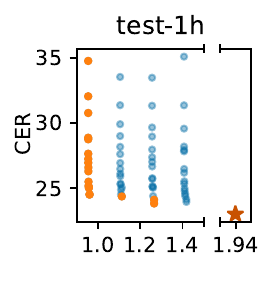}}\\
    \vspace{-0.5em}
    \includegraphics[width=0.4\textwidth]{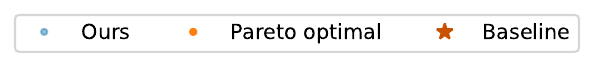}
    \vspace{-0.5em}
    \caption{
    Pareto tradeoff between ASR performance (WER on Librispeech, CER on ML-SUPERB 1h) and computational efficiency (TFLOPs per 1 minute input) for various modifications on WavLM-large.
    %X-axis denotes the computational overhead per generating a single discrete speech unit, measured by GFLOPs.
    %Y-axis denotes the downstream ASR performance, measured by CER on ML-SUPERB 1h test set.
    % Baseline (vanilla WavLM-large) assumes 1 minute of input speech.
    Results show that we can extract DSUs in a lightweight way with limited impact on ASR performance.
    % , while ours enable streaming capabilities.
    }
    \vspace{-1em}
    \label{fig:pareto}
\end{figure}

%% file: figures/fig_window.tex
\begin{figure}[t]
    \centering
    \subfloat{\includegraphics[width=0.15\textwidth]{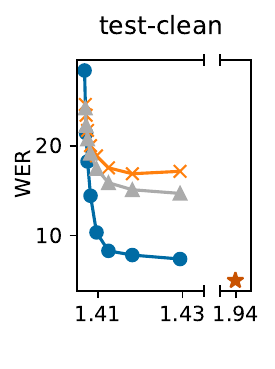} \label{fig:window-clean}}
    \subfloat{\includegraphics[width=0.15\textwidth]{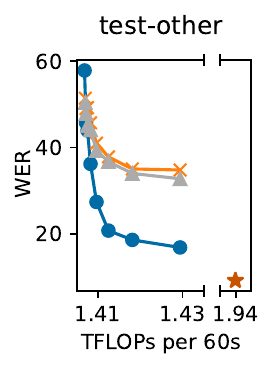} \label{fig:window-other}}
    \subfloat{\includegraphics[width=0.15\textwidth]{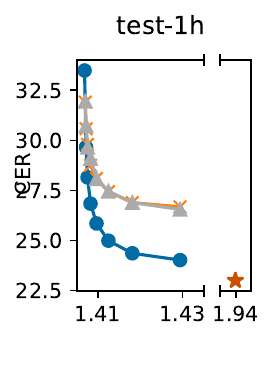} \label{fig:window-1h}}\\
    \vspace{-0.5em}
    \includegraphics[width=0.45\textwidth]{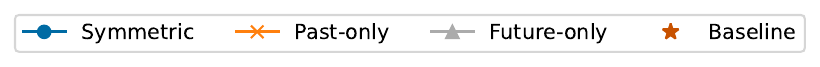}
    \vspace{-0.5em}
    \caption{
        Results on various attention window sizes when extracting DSUs.
        Baseline uses the full speech input, where others use sparse attention with symmetric and asymmetric windows.
    }
    \label{fig:window}
\end{figure}

%% file: figures/fig_baseline_comparison.tex
% Original figure: https://docs.google.com/presentation/d/1QnEX-5Guv9qg1gZxa5FiWkKTYnop6ETvh-9hdNwNrMc/edit?usp=sharing

\begin{figure}[t]
    \centering
    \includegraphics[width=0.47\textwidth]{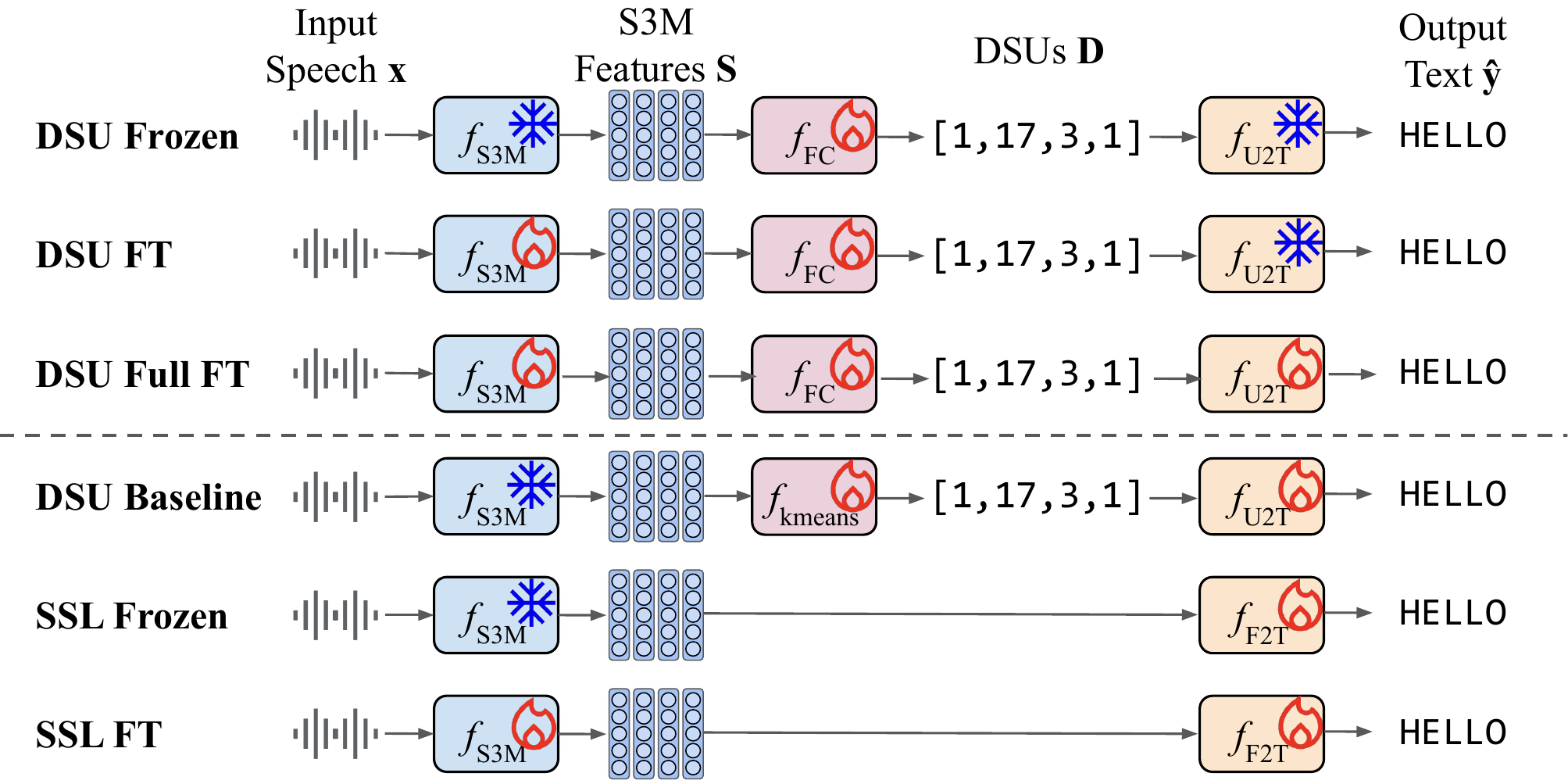}
    \caption{
        Various baselines for \Cref{sec:depth} and \Cref{fig:layer}.
    }
    \vspace{-1em}
    \label{fig:baseline_comp}
\end{figure}

%% file: figures/fig_layer.tex
\begin{figure}[t]
    \centering
    \subfloat{\includegraphics[width=0.15\textwidth]{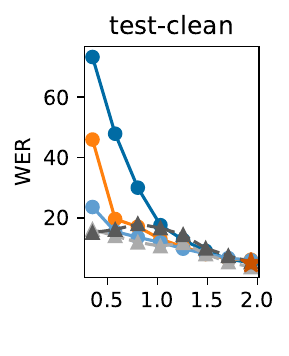} \label{fig:layer-clean}}
    \subfloat{\includegraphics[width=0.15\textwidth]{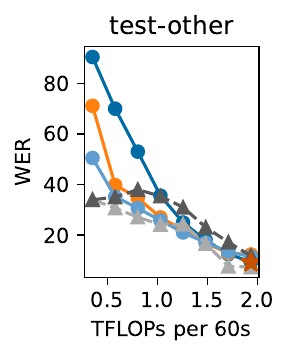} \label{fig:layer-other}}
    \subfloat{\includegraphics[width=0.15\textwidth]{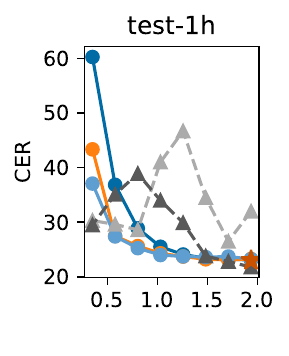} \label{fig:layer-1h}}\\
    \vspace{-0.5em}
    \includegraphics[width=0.37\textwidth]{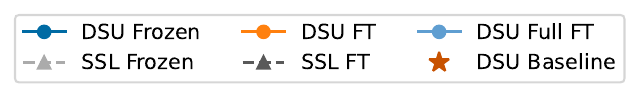}
    \caption{
        Results using features from various layers when extracting DSUs.
        Each methods are summarized in \Cref{fig:baseline_comp}.
    }
    \vspace{-1em}
    \label{fig:layer}
\end{figure}

%% file: figures/fig_sota_window.tex
\begin{figure}[t]
    \centering
    \subfloat{\includegraphics[width=0.15\textwidth]{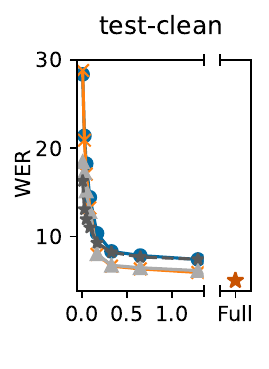}}
    \subfloat{\includegraphics[width=0.15\textwidth]{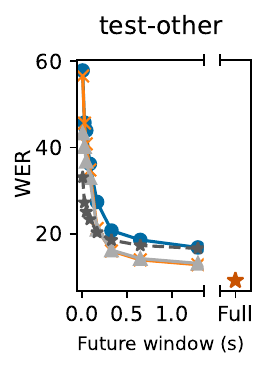}}
    \subfloat{\includegraphics[width=0.15\textwidth]{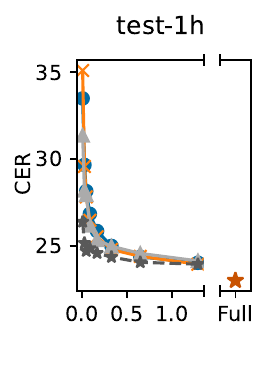}}\\
    \vspace{-0.5em}
    \includegraphics[width=0.45\textwidth]{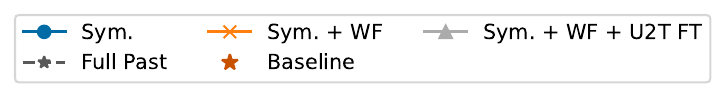}
    \caption{
        Results with varying symmetric attention window sizes (Sym.), with applying additional methods, i.e., learnable weights (WF) and fine-tuning U2T module (U2T FT).
        Baseline uses the full speech input, while Full Past uses full past, but limited future attention window.
    }

    \label{fig:sota-window}
\end{figure}

%% file: figures/appendix_tables.tex
\section{Appendix}

\begin{table}[h]
\caption{We provide the exact values of \Cref{fig:window}.
We denote various attention window configurations as the number of left, center, and right frames, \textit{i.e.}, $[l, c=1, r]$.
As mentioned in \Cref{ssec:setting_detail}, our baseline differs from \cite{chang24b_interspeech} only by the number of beam size, where we reduce from 20 to 5.
As mentioned in \Cref{ssec:setting_eval}, we use WER for test-clean and test-other, and CER for test-1h.
}
\resizebox{1\linewidth}{!}{%
\begin{tabular}{l|rr|rrr|r}
\toprule
\textbf{Method} & $\boldsymbol{l}$ & $\boldsymbol{r}$ & \textbf{test-clean} & \textbf{test-other} & \textbf{test-1h} & \textbf{TFLOPs}\\
\midrule
Baseline \cite{chang24b_interspeech}    &$\infty$&$\infty$                        & 5.0                            & 9.1                            & 22.9                        & 1.936                      \\
Baseline (Ours)                     & $\infty$ &$\infty$ & 5.0                            & 9.2                            & 23.0                        & 1.936                      \\
\midrule
\multirow{8}{*}{Symmetric}   & 0                            & 0                            & 28.37                          & 57.84                          & 33.49                       & 1.407                      \\
                             & 1                            & 1                            & 21.40                           & 45.76                          & 29.66                       & 1.407                      \\
                             & 2                            & 2                            & 18.25                          & 43.96                          & 28.15                       & 1.408                      \\
                             & 4                            & 4                            & 14.43                          & 36.21                          & 26.84                       & 1.408                      \\
                             & 8                            & 8                            & 10.35                          & 27.39                          & 25.85                       & 1.410                      \\
                             & 16                           & 16                           & 8.30                            & 20.76                          & 24.99                       & 1.412                      \\
                             & 32                           & 32                           & 7.83                           & 18.60                           & 24.36                       & 1.418                      \\
                             & 64                           & 64                           & 7.39                           & 16.85                          & 24.02                       & 1.429                      \\
\midrule
\multirow{8}{*}{Past-only}   & 1                            & 0                            & 24.59                          & 51.42                          & 31.91                       & 1.407                      \\
                             & 2                            & 0                            & 23.39                          & 49.00                             & 30.58                       & 1.407                      \\
                             & 4                            & 0                            & 21.70                           & 49.24                          & 29.80                        & 1.408                      \\
                             & 8                            & 0                            & 20.02                          & 45.73                          & 28.75                       & 1.408                      \\
                             & 16                           & 0                            & 18.86                          & 41.03                          & 28.10                        & 1.410                      \\
                             & 32                           & 0                            & 17.54                          & 37.82                          & 27.44                       & 1.412                      \\
                             & 64                           & 0                            & 16.88                          & 35.04                          & 26.89                       & 1.418                      \\
                             & 128                          & 0                            & 17.15                          & 34.79                          & 26.68                       & 1.429                      \\
\midrule
\multirow{8}{*}{Future-only} & 0                            & 1                            & 24.20                           & 50.39                          & 31.95                       & 1.407                      \\
                             & 0                            & 2                            & 22.23                          & 47.84                          & 30.73                       & 1.407                      \\
                             & 0                            & 4                            & 20.73                          & 45.08                          & 29.65                       & 1.408                      \\
                             & 0                            & 8                            & 19.16                          & 44.09                          & 29.09                       & 1.408                      \\
                             & 0                            & 16                           & 17.45                          & 39.29                          & 28.08                       & 1.410                      \\
                             & 0                            & 32                           & 15.88                          & 36.72                          & 27.46                       & 1.412                      \\
                             & 0                            & 64                           & 15.10                           & 33.99                          & 26.89                       & 1.418                      \\
                             & 0                            & 128                          & 14.69                          & 32.77                          & 26.55                       & 1.429                     \\
\bottomrule
\end{tabular}
}%
\end{table}

\begin{table}[ht]
\caption{
We provide the exact values of \Cref{fig:layer}.
Layers denote the number of layers used.
As mentioned in \Cref{ssec:setting_eval}, we use WER for test-clean and test-other, and CER for test-1h.
For SSL Frozen and SSL FT, we modified the configuration of the existing baseline.
}
\resizebox{1\linewidth}{!}{%
\begin{tabular}{l|r|rrr|r}
\toprule
\textbf{Method} & \textbf{Layer} & \textbf{test-clean} & \textbf{test-other} & \textbf{test-1h} & \textbf{TFLOPs}\\
\midrule
Baseline              & 21                        & 5.0                            & 9.2                            & 23.0                        & 1.936                      \\
\midrule 
\multirow{8}{*}{DSU Frozen}  & 21                        & 5.11                           & 9.41                           & 23.08                       & 1.936                      \\
                             & 18                        & 6.30                           & 12.51                          & 23.23                       & 1.709                      \\
                             & 15                        & 9.03                           & 17.98                          & 23.40                       & 1.483                      \\
                             & 12                        & 12.67                          & 24.99                          & 24.09                       & 1.256                      \\
                             & 9                         & 17.56                          & 35.60                          & 25.51                       & 1.029                      \\
                             & 6                         & 29.99                          & 53.02                          & 28.89                       & 0.802                      \\
                             & 3                         & 47.87                          & 69.99                          & 36.85                       & 0.576                      \\
                             & 0                         & 73.29                          & 90.42                          & 60.26                       & 0.349                      \\
\midrule
\multirow{8}{*}{DSU FT}      & 21                        & 6.11                           & 12.35                          & 23.26                       & 1.936                      \\
                             & 18                        & 6.34                           & 12.85                          & 23.10                       & 1.709                      \\
                             & 15                        & 8.00                           & 16.88                          & 23.21                       & 1.483                      \\
                             & 12                        & 10.37                          & 22.06                          & 23.72                       & 1.256                      \\
                             & 9                         & 12.96                          & 26.99                          & 24.31                       & 1.029                      \\
                             & 6                         & 17.05                          & 34.70                          & 25.63                       & 0.802                      \\
                             & 3                         & 19.60                          & 39.80                          & 27.48                       & 0.576                      \\
                             & 0                         & 45.94                          & 71.13                          & 43.35                       & 0.349                      \\
\midrule
\multirow{8}{*}{DSU Full FT} & 21                        & 5.94                           & 11.48                          & 23.25                       & 1.936                      \\
                             & 18                        & 6.65                           & 13.40                          & 23.72                       & 1.709                      \\
                             & 15                        & 8.09                           & 17.02                          & 23.61                       & 1.483                      \\
                             & 12                        & 9.77                           & 21.05                          & 23.73                       & 1.256                      \\
                             & 9                         & 11.85                          & 25.79                          & 23.99                       & 1.029                      \\
                             & 6                         & 13.56                          & 30.68                          & 25.25                       & 0.802                      \\
                             & 3                         & 15.46                          & 35.16                          & 27.40                       & 0.576                      \\
                             & 0                         & 23.58                          & 50.51                          & 37.08                       & 0.349                      \\
\midrule
\multirow{8}{*}{SSL Frozen}  & 21             & 5.3                 & 11.8                & 21.8             & 1.936           \\
                             & 18             & 7.4                 & 17.2                & 22.8             & 1.709           \\
                             & 15             & 9.9                 & 23.1                & 23.8             & 1.483           \\
                             & 12             & 14.4                & 31.0                & 29.9             & 1.256           \\
                             & 9              & 16.6                & 35.4                & 34.0             & 1.029           \\
                             & 6              & 18.0                & 38.0                & 38.9             & 0.802           \\
                             & 3              & 16.1                & 35.0                & 35.1             & 0.576           \\
                             & 0              & 15.1                & 34.0                & 29.5             & 0.349           \\
\midrule
\multirow{8}{*}{SSL FT}      & 21             & 3.7                 & 7.4                 & 32.0             & 1.936           \\
                             & 18             & 5.3                 & 7.7                 & 26.5             & 1.709           \\
                             & 15             & 8.0                 & 16.7                & 34.5             & 1.483           \\
                             & 12             & 11.7                & 24.3                & 46.7             & 1.256           \\
                             & 9              & 10.6                & 24.2                & 41.0             & 1.029           \\
                             & 6              & 11.9                & 27.1                & 28.6             & 0.802           \\
                             & 3              & 14.0                & 30.7                & 29.6             & 0.576           \\
                             & 0              & 16.1                & 34.0                & 30.3             & 0.349           \\ 
                             \bottomrule
\end{tabular}
}%
\end{table}

\begin{table}[ht]
\caption{We provide the exact values of \Cref{fig:sota-window}.
We denote various attention window configurations as the number of left, center, and right frames, \textit{i.e.}, $[l, c=1, r]$.
Sym., WF, and U2T FT denotes symmetric attention window, learnable weights, and fine-tuning U2T module, respectively.
Baseline uses the full speech input, while Full Past uses full past, but limited future attention window.
As mentioned in \Cref{ssec:setting_eval}, we use WER for test-clean and test-other, and CER for test-1h.
}
\resizebox{1\linewidth}{!}{%
\begin{tabular}{l|rr|rrr|r}
\toprule
\textbf{Method} & $\boldsymbol{l}$ & $\boldsymbol{r}$ & \textbf{test-clean} & \textbf{test-other} & \textbf{test-1h} & \textbf{TFLOPs}\\
\midrule
Baseline                        & $\infty$ & $\infty$ & 5.0                  & 9.2                  & 23.0                 & 1.936                \\
\midrule \multirow{8}{*}{Sym.}           & 0                            & 0                            & 28.37                & 57.84                & 33.49                & 1.407                \\
                                & 1                            & 1                            & 21.40                 & 45.76                & 29.66                & 1.407                \\
                                & 2                            & 2                            & 18.25                & 43.96                & 28.15                & 1.408                \\
                                & 4                            & 4                            & 14.43                & 36.21                & 26.84                & 1.408                \\
                                & 8                            & 8                            & 10.35                & 27.39                & 25.85                & 1.410                \\
                                & 16                           & 16                           & 8.30                  & 20.76                & 24.99                & 1.412                \\
                                & 32                           & 32                           & 7.83                 & 18.60                 & 24.36                & 1.418                \\
                                & 64                           & 64                           & 7.39                 & 16.85                & 24.02                & 1.429                \\
\midrule \multirow{8}{*}{Sym.+WF}        & 0                            & 0                            & 28.81                & 56.55                & 35.10                 & 1.407                \\
                                & 1                            & 1                            & 20.91                & 45.66                & 29.57                & 1.407                \\
                                & 2                            & 2                            & 17.07                & 40.83                & 27.82                & 1.407                \\
                                & 4                            & 4                            & 13.23                & 34.81                & 26.47                & 1.408                \\
                                & 8                            & 8                            & 7.96                 & 21.24                & 25.52                & 1.408                \\
                                & 16                           & 16                           & 6.63                 & 16.04                & 24.79                & 1.410                \\
                                & 32                           & 32                           & 6.32                 & 13.97                & 24.39                & 1.412                \\
                                & 64                           & 64                           & 5.91                 & 12.86                & 23.94                & 1.418                \\
\midrule \multirow{8}{*}{Sym.+WF+U2T FT} & 0                            & 0                            & 18.45                & 43.36                & 31.36                & 1.407                \\
                                & 1                            & 1                            & 17.23                & 40.24                & 28.19                & 1.407                \\
                                & 2                            & 2                            & 15.15                & 36.79                & 27.93                & 1.407                \\
                                & 4                            & 4                            & 12.82                & 33.04                & 26.18                & 1.408                \\
                                & 8                            & 8                            & 8.02                 & 21.03                & 25.34                & 1.408                \\
                                & 16                           & 16                           & 6.72                 & 16.24                & 24.97                & 1.410                \\
                                & 32                           & 32                           & 6.45                 & 14.28                & 24.56                & 1.412                \\
                                & 64                           & 64                           & 6.12                 & 13.17                & 24.13                & 1.418                \\
\midrule \multirow{8}{*}{Full Past}      & $\infty$ & 0                            & 16.28                & 33.12                & 26.38                & 1.671                \\
                                & $\infty$ & 1                            & 13.08                & 27.26                & 25.16                & 1.672                \\
                                & $\infty$ & 2                            & 11.94                & 25.09                & 24.70                 & 1.672                \\
                                & $\infty$ & 4                            & 11.01                & 23.35                & 24.85                & 1.672                \\
                                & $\infty$ & 8                            & 9.27                 & 20.34                & 24.58                & 1.672                \\
                                & $\infty$ & 16                           & 8.20                  & 18.55                & 24.35                & 1.673                \\
                                & $\infty$ & 32                           & 7.67                 & 17.32                & 24.05                & 1.674                \\
                                & $\infty$ & 64                           & 7.41                 & 16.62                & 23.94                & 1.677          \\ \bottomrule     
\end{tabular}
}%
\end{table}

\begin{table}[ht]
\caption{
We provide the exact values for each individual points of \Cref{fig:pareto}.
We apply both symmetric window and learnable weights (Sym. + WF) while varying the window size and the number of layers.
As mentioned in \Cref{ssec:setting_eval}, we use WER for test-clean and test-other, and CER for test-1h.
}
\resizebox{1\linewidth}{!}{%
\begin{tabular}{l|rr|rrr|r}
\toprule
\textbf{Layer} & $\boldsymbol{l}$ & $\boldsymbol{r}$ & \textbf{test-clean} & \textbf{test-other} & \textbf{test-1h} & \textbf{TFLOPs}\\
\midrule \multirow{8}{*}{21} & 0                    & 0                    & 28.81                & 56.55                & 35.10                 & 1.407                \\
                    & 1                    & 1                    & 20.91                & 45.66                & 29.57                & 1.407                \\
                    & 2                    & 2                    & 17.07                & 40.83                & 27.82                & 1.407                \\
                    & 4                    & 4                    & 13.23                & 34.81                & 26.47                & 1.408                \\
                    & 8                    & 8                    & 7.96                 & 21.24                & 25.52                & 1.408                \\
                    & 16                   & 16                   & 6.63                 & 16.04                & 24.79                & 1.410                \\
                    & 32                   & 32                   & 6.32                 & 13.97                & 24.39                & 1.412                \\
                    & 64                   & 64                   & 5.91                 & 12.86                & 23.94                & 1.418                \\
\midrule \multirow{8}{*}{18} & 0                    & 0                    & 28.45                & 57.28                & 33.48                & 1.256                \\
                    & 1                    & 1                    & 22.37                & 48.00                   & 29.66                & 1.256                \\
                    & 2                    & 2                    & 18.21                & 43.61                & 27.96                & 1.256                \\
                    & 4                    & 4                    & 14.36                & 35.94                & 26.66                & 1.256                \\
                    & 8                    & 8                    & 9.32                 & 24.81                & 25.70                 & 1.257                \\
                    & 16                   & 16                   & 8.09                 & 19.52                & 25.09                & 1.258                \\
                    & 32                   & 32                   & 7.39                 & 17.59                & 24.39                & 1.261                \\
                    & 64                   & 64                   & 6.92                 & 15.71                & 24.10                 & 1.265                \\
\midrule \multirow{8}{*}{15} & 0                    & 0                    & 28.63                & 57.35                & 33.54                & 1.105                \\
                    & 1                    & 1                    & 23.09                & 49.32                & 29.91                & 1.105                \\
                    & 2                    & 2                    & 18.52                & 43.81                & 28.17                & 1.105                \\
                    & 4                    & 4                    & 15.30                 & 38.43                & 26.93                & 1.105                \\
                    & 8                    & 8                    & 11.90                 & 29.92                & 25.91                & 1.106                \\
                    & 16                   & 16                   & 10.62                & 25.09                & 25.28                & 1.107                \\
                    & 32                   & 32                   & 9.90                  & 22.18                & 24.52                & 1.109                \\
                    & 64                   & 64                   & 9.43                 & 20.82                & 24.36                & 1.113                \\
\midrule \multirow{8}{*}{12} & 0                    & 0                    & 29.32                & 57.95                & 34.77                & 0.953                \\
                    & 1                    & 1                    & 23.82                & 48.84                & 30.77                & 0.954                \\
                    & 2                    & 2                    & 20.27                & 47.22                & 28.75                & 0.954                \\
                    & 4                    & 4                    & 16.55                & 40.36                & 27.24                & 0.954                \\
                    & 8                    & 8                    & 14.18                & 34.35                & 26.29                & 0.954                \\
                    & 16                   & 16                   & 13.04                & 30.58                & 25.49                & 0.955                \\
                    & 32                   & 32                   & 12.36                & 27.50                 & 25.02                & 0.957                \\
                    & 64                   & 64                   & 11.92                & 26.21                & 24.52                & 0.960               \\
\bottomrule
\end{tabular}
}%
\end{table}

\begin{table}[ht]
\caption{
We provide the exact values for each individual points of \Cref{fig:pareto}.
We apply all symmetric window, learnable weights, and fine-tuning U2T module (Sym. + WF + U2T FT) while varying the window size and the number of layers.
As mentioned in \Cref{ssec:setting_eval}, we use WER for test-clean and test-other, and CER for test-1h.
}
\resizebox{1\linewidth}{!}{%
\begin{tabular}{l|rr|rrr|r}
\toprule
\textbf{Layer} & $\boldsymbol{l}$ & $\boldsymbol{r}$ & \textbf{test-clean} & \textbf{test-other} & \textbf{test-1h} & \textbf{TFLOPs}\\
\midrule \multirow{8}{*}{21} & 0                    & 0                    & 18.45                & 43.36                & 31.36                & 1.407                \\
                    & 1                    & 1                    & 17.23                & 40.24                & 28.19                & 1.407                \\
                    & 2                    & 2                    & 15.15                & 36.79                & 27.93                & 1.407                \\
                    & 4                    & 4                    & 12.82                & 33.04                & 26.18                & 1.408                \\
                    & 8                    & 8                    & 8.02                 & 21.03                & 25.34                & 1.408                \\
                    & 16                   & 16                   & 6.72                 & 16.24                & 24.97                & 1.410                \\
                    & 32                   & 32                   & 6.45                 & 14.28                & 24.56                & 1.412                \\
                    & 64                   & 64                   & 6.12                 & 13.17                & 24.13                & 1.418                \\
\midrule \multirow{8}{*}{18} & 0                    & 0                    & 19.34                & 44.96                & 31.33                & 1.256                \\
                    & 1                    & 1                    & 17.11                & 40.86                & 28.96                & 1.256                \\
                    & 2                    & 2                    & 15.18                & 37.84                & 27.37                & 1.256                \\
                    & 4                    & 4                    & 14.00                & 33.75                & 27.02                & 1.256                \\
                    & 8                    & 8                    & 9.37                 & 24.83                & 25.76                & 1.257                \\
                    & 16                   & 16                   & 8.08                 & 19.45                & 25.15                & 1.258                \\
                    & 32                   & 32                   & 7.76                 & 17.85                & 25.02                & 1.261                \\
                    & 64                   & 64                   & 6.97                 & 15.88                & 23.82                & 1.265                \\
\midrule \multirow{8}{*}{15} & 0                    & 0                    & 19.70                & 45.20                & 31.37                & 1.105                \\
                    & 1                    & 1                    & 17.48                & 40.70                & 29.00                & 1.105                \\
                    & 2                    & 2                    & 15.57                & 37.85                & 27.61                & 1.105                \\
                    & 4                    & 4                    & 14.17                & 34.81                & 26.47                & 1.105                \\
                    & 8                    & 8                    & 11.98                & 29.16                & 26.10                & 1.106                \\
                    & 16                   & 16                   & 10.21                & 24.65                & 25.35                & 1.107                \\
                    & 32                   & 32                   & 9.91                 & 22.18                & 24.87                & 1.109                \\
                    & 64                   & 64                   & 9.60                 & 20.58                & 25.04                & 1.113                \\
\midrule \multirow{8}{*}{12} & 0                    & 0                    & 19.01                & 44.19                & 32.05                & 0.953                \\
                    & 1                    & 1                    & 17.42                & 40.54                & 28.87                & 0.954                \\
                    & 2                    & 2                    & 15.69                & 38.70                & 27.64                & 0.954                \\
                    & 4                    & 4                    & 14.60                & 35.54                & 26.95                & 0.954                \\
                    & 8                    & 8                    & 13.54                & 32.64                & 26.58                & 0.954                \\
                    & 16                   & 16                   & 12.43                & 29.28                & 25.17                & 0.955                \\
                    & 32                   & 32                   & 11.79                & 27.19                & 25.05                & 0.957                \\
                    & 64                   & 64                   & 11.50                & 25.63                & 24.50                & 0.960   \\            
\bottomrule
\end{tabular}
}%
\end{table}